\documentclass[sigconf, authorversion]{acmart}

\AtBeginDocument{%
  }

\copyrightyear{2026}
\acmYear{2026}
\setcopyright{cc}
\setcctype{by}
\acmConference[ETRA '26]{2026 Symposium on Eye Tracking Research and Applications}{June 01--04, 2026}{Marrakesh, Morocco}
\acmBooktitle{2026 Symposium on Eye Tracking Research and Applications (ETRA '26), June 01--04, 2026, Marrakesh, Morocco}
\acmDOI{10.1145/3797246.3806212}
\acmISBN{979-8-4007-2519-7/2026/06}
  \acmConference[ETRA 2026]{ ACM Symposium on Eye Tracking Research \& Applications}{June 1--14,
  2026}{ Marrakech, Morocco}
\acmPrice{15.00}
\acmISBN{978-1-4503-XXXX-X/18/06}

\acmSubmissionID{1013}

\begin{document}

\title[As Far as Eye see]{As Far as Eye See: Vergence-Pupil Coupling in Near-Far Depth Switching}

\author{Virmarie Maquiling}
\email{virmarie.maquiling@tum.de}
\affiliation{%
	\institution{Human-Centered Technologies for Learning, Technical University of Munich}
	\streetaddress{Marsstraße 20-22}
	\city{Munich} 
	\country{Germany} 
	\postcode{80335}
}
\affiliation{%
  \institution{Munich Center for Machine Learning (MCML)}
  \city{Munich}
  \country{Germany}
}

\author{Yasmeen Abdrabou}
\email{yasmeen.abdrabou@tum.de}
\affiliation{%
	\institution{Human-Centered Technologies for Learning, Technical University of Munich}
	\streetaddress{Marsstraße 20-22}
	\city{Munich} 
	\country{Germany} 
	\postcode{80335}
}
\affiliation{%
  \institution{Munich Center for Machine Learning (MCML)}
  \city{Munich}
  \country{Germany}
}

\author{Enkelejda Kasneci}
\email{enkelejda.kasneci@tum.de}
\affiliation{%
	\institution{Human-Centered Technologies for Learning, Technical University of Munich}
	\streetaddress{Marsstraße 20-22}
	\city{Munich} 
	\country{Germany} 
	\postcode{80335}
}
\affiliation{%
  \institution{Munich Center for Machine Learning (MCML)}
  \city{Munich}
  \country{Germany}
}

\renewcommand{\shortauthors}{Maquiling et al.}

\begin{abstract} % max 150: 150
Vergence is widely used as a proxy for depth perception and spatial attention in immersive and real-world eye-tracking studies. In this paper, we investigate how pupil size artefacts affect vergence estimates during real physical depth viewing with a head-mounted eye tracker. Using a beamsplitter setup with physically near and far targets, we elicited controlled convergent and divergent eye movements under static, luminance-modulated, and blockwise fixation conditions. Near and far targets were reliably separable in vergence angle across participants. However, pupil-vergence coupling varied substantially across individuals and conditions. Static illumination produced large inter-participant variability, while luminance modulation reduced this spread, yielding more clustered estimates. Blockwise and audio-cued recordings further showed that pupil-vergence coupling persists even without visual depth onsets. These results suggest that pupil size fluctuations can systematically influence vergence estimates, and that controlled viewing conditions can reduce—but not eliminate—this effect.
\end{abstract}



% \keywords{Vergence, Pupillometry, Eye tracking, Pupil Size Artefacts, Depth-switching}
\begin{teaserfigure}
    \centering
  \includegraphics[width=.7\textwidth]{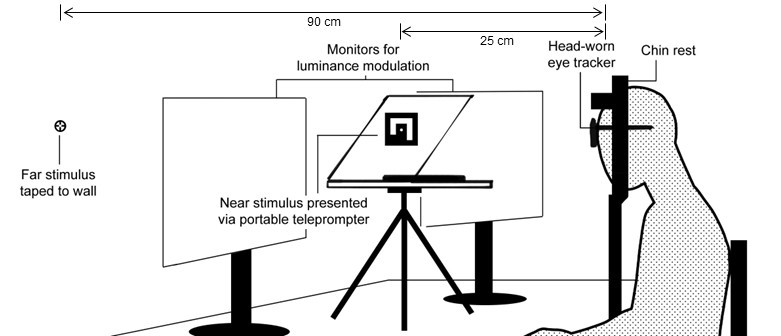} 
  \caption{In this experiment, the participants were instructed to alternate their fixation between a target projected through a beamsplitter and another target pasted to a wall. The two targets were positioned roughly along the participant's viewing axis to minimize unnecessary vertical and horizontal gaze rotations.} 
  \label{fig:teaser}
\end{teaserfigure}

% \received{20 January 2026}
% \received[revised]{--}
% \received[accepted]{--}

\maketitle

\section{Introduction}

Depth perception is fundamental to interaction in 3D environments. In everyday tasks, such as a cook grabbing ingredients from a nearby countertop or a driver reading the speedometer on their dashboard while maintaining awareness of the road, humans continuously rely on accurate depth information. In immersive virtual and augmented reality (VR/AR) systems, depth plays an equally critical role. Techniques such as foveated rendering~\cite{guenter2012foveated, wang2023foveated}, gaze-contingent depth-of-field effects~\cite{mauderer2014depth,duchowski2014reducing, mantiuk2011gaze}, and gaze-based selection~\cite{schweigert2019eyepointing, kumar2007eyepoint} all assume that a system can reliably infer where in depth a user is looking. When this assumption fails--by sharpening or highlighting the wrong object--immersion can break, discomfort can increase, and the user can experience VR sickness. As a result, vergence is often regarded as a promising cue for estimating gaze depth in eye-tracked XR systems. 

Many gaze-based depth estimation approaches start from a geometric relationship between vergence and depth: if the orientations of both eyes are known, viewing distance can be recovered by finding the intersection between the two lines of sight in 3D space~\cite{collewijn1997trajectories}. In practice, however, video-based eye tracking (VOG) does not measure eyeball rotation directly. Instead, gaze direction is inferred from image-based features such as the pupil center~\cite{hansen2009eye}, making the estimated eye orientation inherently sensitive to pupil dynamics. %
The pupil is not a rigid marker but a dynamic aperture whose apparent position shifts with dilation and constriction, inducing apparent changes in the estimated pupil center that can be misinterpreted as eye rotation. This pupil size artefact (PSA) has been shown to systematically bias gaze and vergence estimates in video-based eye tracking, even when the eyeball itself remains stationary~\cite{hooge2019pupil, drewes2014smaller, jaschinski2016pupil}. At the same time, vergence, accommodation, and pupil size are physiologically coordinated as part of the near triad during changes in viewing distance~\cite{myers1990topology}, yet this coupling is rarely accounted for in applied vergence-based depth estimation.

Existing work has largely minimized pupil dynamics, treated pupil size as a luminance or cognitive signal, or examined vergence under stereo disparity rather than real physical depth. What remains unclear is whether the structure of vergence error itself changes with physical viewing state, rather than behaving as a fixed bias. We therefore ask whether pupil-vergence coupling manifests during physically grounded near-far viewing, whether its strength and direction depend on viewing depth and task, and whether constraining pupil dynamics through luminance modulation reduces inter-participant variability. %
We address these questions using a beamsplitter-based setup that presents physically near and far targets along the viewing axis, enabling controlled vergence under real depth rather than simulated disparity. Across near-far switching, static and luminance-modulated viewing, and sustained fixation, we analyze the relationship between vergence and pupil size. We show that near and far depth states are reliably separable via vergence, but that pupil-vergence coupling varies across participants and conditions and persists even under constrained viewing, suggesting that pupil size is a systematic confound rather than mere noise in vergence-based depth estimation.

This work examines the validity of vergence as a depth-related measurement in video-based eye tracking and suggests that pupil size artefacts introduce structured, state-dependent distortions rather than random noise. As a result, vergence estimates are not purely geometric but depend on viewing conditions and pupil dynamics, and must therefore be interpreted accordingly.

\section{Related Works} 

\subsection{Pupil-Vergence Coupling and the Near Triad}
The coordinated involvement of vergence, accommodation, and pupil size during changes in viewing distance--commonly referred to as the near triad--has long been described in oculomotor research~\cite{myers1990topology, hung1984near}. Rather than being solely driven by luminance, pupil responses have been shown to interact with vergence as part of depth-related oculomotor behavior. In particular, dynamic pupil-vergence coupling has been observed during disparity-driven vergence responses, with coupling strength varying across task phases and, in some cases, reversing sign within participants~\cite{balaban2018patterns}. These findings suggest that pupil-vergence coupling is state-dependent rather than governed by a single fixed relationship. Complementary evidence from studies using artificial pupils shows that pupil size systematically modulates the accommodation-vergence relationship by affecting measured AC/A ratios at very small pupil sizes even under controlled luminance~\cite{ripps1962effect}. At the same time, pupil size is known to be modulated by attentional, cognitive and arousal-related factors~\cite{mathot2015new,mathot2018pupillometry, kahnemann1966pupil}, further complicating the interpretation of pupil-linked effects in vergence analyses.

\subsection{Pupil Size and Vergence-Based Gaze Depth Estimation}
Pupil size artefacts have been an established source of bias in video-based eye tracking. \citeauthor{drewes2014smaller}~[\citeyear{drewes2014smaller}] showed systematic gaze shifts with changing pupil diameter, and subsequent work demonstrated that vergence estimates vary strongly with pupil size even during fixation at a single depth, such that pupil-related effects can masquerade as depth changes and limit purely geometric interpretations of vergence~\cite{hooge2019pupil}. Pupil-related bias has further been shown to persist under controlled fixation and to depend on viewing distance, particularly in near viewing~\cite{jaschinski2016pupil}. Vergence has nevertheless long been used to estimate gaze depth, with early work demonstrating that binocular geometry can recover fixation distance under controlled conditions~\cite{duchowski2011measuring}, but later studies revealing substantial sensitivity to calibration distance, measurement error, and viewing setup~\cite{duchowski2014comparing}. Despite these limitations, vergence estimates from video-based eye trackers remain sufficient for distinguishing coarse depth states, although this separability does not imply accurate absolute vergence or a purely geometric depth signal. To improve robustness, subsequent approaches combine vergence with additional signals or learning-based mappings (e.g., \citeauthor{lee2017estimating}~\citeyear{lee2017estimating}, \citeauthor{oney2020evaluation}~\citeyear{oney2020evaluation}, \citeauthor{von2025cnn}~\citeyear{von2025cnn}, \citeauthor{cho2024hybrid}~\citeyear{cho2024hybrid}). While these methods reduce variance and extend usable depth ranges, they typically treat vergence as a geometric signal corrupted by noise, with pupil-vergence coupling and its state dependence rarely modeled explicitly. In practice, pupil-related distortions are absorbed into model parameters, limiting interpretability across tasks, illumination conditions, and viewing states.

Despite this body of work, it is still unclear whether pupil-vergence coupling in video-based eye tracking can be treated as a fixed, correctable bias during physically grounded, near-far viewing. Existing PSA-focused studies typically either examine vergence under controlled luminance during static fixation at a single depth or under stereo disparity. Meanwhile, applied gaze-depth estimation methods typically address vergence error through calibration or learned mappings, but rarely test whether pupil-vergence coupling introduces state-dependent error across viewing distance, illumination, and task context~\cite{lee2017estimating, duchowski2011measuring, duchowski2014comparing, oney2020evaluation, cho2024hybrid}

\section{Methods}

\subsection{Study Design}
The study used a within-subjects design in which each participant completed multiple controlled viewing tasks involving physically grounded near and far fixation. The independent variables were viewing distance (near/far), illumination condition (static vs modulated luminance), and task (near-far switching, blockwise fixation in near/far depth, audio-cued control). The dependent variables were vergence and PSA slope (see definition below). Analysis were conducted primarily within participants, with each participant serving as their own control across tasks and conditions. 

\subsection{Apparatus and Setup}

Participants were seated with head stabilization provided by a chin rest and wore Pupil Labs Neon eye-tracking glasses. Neon is an IR-illuminated, video-based eye tracker that estimates gaze using a proprietary, end-to-end deep learning pipeline ("NeonNet") operating on concurrent binocular infrared eye images, without user-specific calibration~\cite{baumann2023neon}. Architectural and feature-level details of the gaze estimation model are not publicly disclosed, including how pupil size variations are handled internally. A near stimulus was presented via a smartphone mounted inside a portable teleprompter using a partially reflective beamsplitter mirror, allowing the near stimulus (an ArUco marker with a white dot in the center) to be optically superimposed into the participant’s view while maintaining visibility of a far fixation target. The far stimulus consisted of an ABC-like fixation target~\cite{thaler2013best} taped to the wall opposite to the participant. The near stimulus was presented for 3\,s while visible and for another 3\,s while not visible. Based on the calibrated scene camera, the projected near ArUco marker was located at approximately 0.25\,m from the participant, while the far marker was positioned at around 0.9\,m. Two monitors placed symmetrically on either side of the participant at a distance of about 0.5\,m either modulated background luminance from black to white at 20 Hz or remained at constant black, depending on the task. The experiment was conducted in a dimly lit room with diffuse natural light. For a visualization of the setup, see Fig.~\ref{fig:teaser}.

\subsection{Tasks and Procedure}
Our goal was not to estimate population-level effect sizes, but to examine whether vergence-pupil coupling is present and how it varies across individuals during controlled near-far viewing. Analyses were therefore conducted primarily within participants, with each participant serving as their own control across tasks. Prior collection, near-far alignment was verified for each participant. Eight participants (six of them right-eye-dominant) with normal to corrected vision completed between four and five tasks. Three participants completed four tasks, while five completed an additional audio-cued control task. The tasks were:

\begin{itemize}
    \item \textbf{Near-far static}: Participants fixated the near ArUco marker when present and shifted fixation to the far target when it disappeared; background luminance was static.
    \item \textbf{Near-far modulated}: Identical to the near-far static task, but with modulated background luminance.
    \item \textbf{Near-far static audio-cued}: Background luminance was static and the near ArUco marker remained continuously visible. Participants switched fixation between near and far targets in response to an auditory cue.
    \item \textbf{PSA near}: With modulated background luminance, participants fixated the near ArUco marker for 30\,s.
    \item \textbf{PSA far}: With modulated background luminance, participants fixated the far target for 30\,s.
\end{itemize}

\subsection{Stimulus Timing and Preprocessing}
Stimulus timing and near target depth were inferred from the calibrated scene camera, while the far target depth was measured by hand. Because background luminance modulation and camera exposure occasionally interfered with ArUco marker detection, we used a manually estimated target identifier reflecting the known timing of near and far stimulus presentation. To account for the slower dynamics of vergence eye movements and transient instability following depth changes~\cite{robinson1966mechanics, hung1998dynamic}, the first 1\,s following each near-far transition was excluded from analysis.

\subsection{Vergence and Pupil Analysis}
Vergence was computed as the angular difference between the normalized binocular gaze vectors and expressed in degrees. Pupil size was estimated as the mean of the left and right pupil diameters. Samples with pupil diameters below 2.2\,mm were excluded to remove blink-related artefacts, and vergence outliers were removed using an interquartile-range criterion (2~IQR). To characterize the relationship between vergence and pupil size within a given fixation depth, vergence-pupil coupling was quantified as the slope of a linear regression between vergence angle and pupil size:
\begin{equation}
\beta_{\mathrm{VP}} = 
\frac{\sum_{i=1}^{N} \left( P_i - \bar{P} \right)\left( V_i - \bar{V} \right)}
{\sum_{i=1}^{N} \left( P_i - \bar{P} \right)^2}
\end{equation}

This slope served as a summary measure of how vergence changed relative to pupil size during sustained fixation. Because our analysis focused on within-epoch vergence-pupil coupling rather than task-evoked pupil responses, we did not apply baseline pupil correction.

\section{Results}

\begin{figure*}[t]
    \centering
    \includegraphics[width=\textwidth]{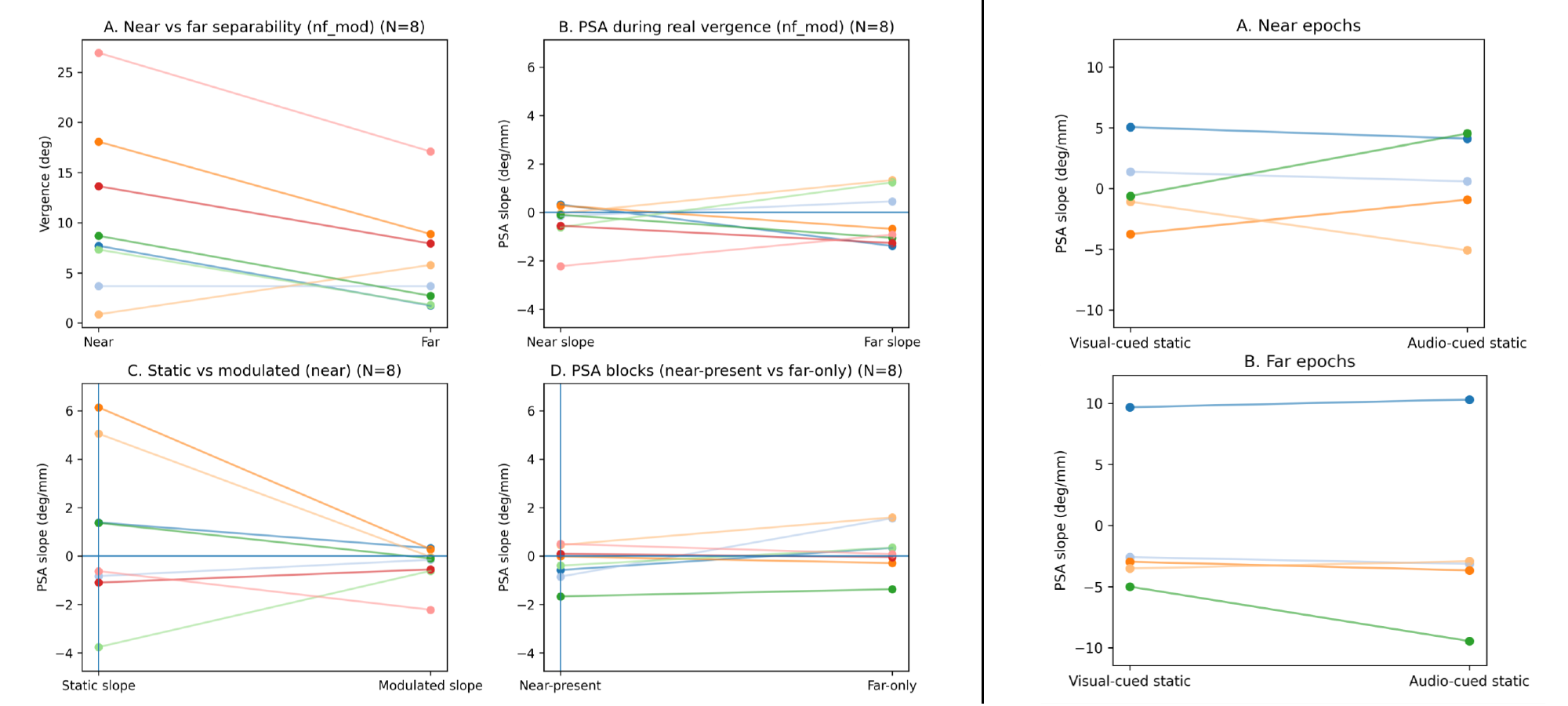}
    \caption{Left: Group-level summary of vergence and pupil-vergence coupling (PSA). Each color represents one participant. Right: (Control) PSA slopes during visual-cued and audio-cued near/far depth-switching tasks with static background illumination. Each color represents one participant.}
    \label{fig:groupsum}
\end{figure*}

Results are analyzed at the participant level, focusing on how pupil–vergence coupling varies across conditions rather than on population-level estimates. The left side of Figure~\ref{fig:groupsum} summarizes vergence behavior and pupil-vergence coupling across participants. Panel~A shows median vergence during near and far epochs in the luminance-modulated near-far task. For most participants, vergence was higher (i.e., more converged) during near viewing than during far viewing. Two participants showed the opposite ordering. However, even in these cases, near and far epochs remained clearly separable within participant. Absolute vergence magnitude varied substantially between individuals. %
Having established reliable separation of depth states, Panel~B examines pupil-vergence coupling (PSA; vergence as a function of pupil size) during depth switching. PSA slopes varied widely in both sign and magnitude across participants and often differed between near and far viewing within the same participant. This suggests that pupil-vergence coupling varies across participants and appears to differ between viewing distance. %
Panel~C focuses on how background illumination affects pupil-vergence coupling during near viewing. Under static illumination, PSA slopes exhibited a wide spread (approximately $-4$ to $+6$~deg/mm). In contrast, luminance modulation reduced this spread, with slopes clustering close to zero. Because the near target was present in both conditions, this reduction is unlikely to be explained by target appearance alone. Instead, luminance modulation was associated with more consistent PSA slope estimates across participants. Although not shown in the figure, the same pattern was observed during far viewing, with greater inter-participant variability under static illumination than under modulated illumination. %
Panel~D examines pupil-vergence coupling during sustained fixation without depth switching. PSA slopes measured during near-only and far-only viewing under luminance-modulated conditions differed between depth states, indicating that pupil-vergence coupling persists during static fixation and depends on the current viewing depth rather than requiring explicit depth transitions. %
The right side of Figure~\ref{fig:groupsum} shows results from an audio-cued control task conducted with a subset of participants. In this condition, the near target remained continuously visible and background illumination was held constant, while depth switching was cued by sound. PSA slopes were again observed for both near and far fixation and differed between visually cued and audio-cued conditions. Pupil-vergence coupling was also observed when depth switching was driven purely by instruction, suggesting that the effect does not rely on visual onset or offset cues.

When considered jointly, these results characterize how vergence and pupil-vergence coupling vary across participants, depth states, and task constraints under physically grounded viewing.

\section{Discussion}
In this work, we set out to examine how pupil size fluctuations influence vergence estimates from video-based eye trackers during real, physically grounded depth viewing. Across tasks and control conditions, pupil-vergence coupling was consistently present, but varied substantially across participants, viewing depths, and task contexts. This variability challenges the common assumption that pupil-related effects on vergence can be treated as negligible or constant, and motivates a closer look at how pupil-vergence coupling changes across viewing states. 

% \vspace{0.5em}

\noindent\textbf{Pupil-vergence coupling is state-dependent, not a fixed bias.}
One of the clearest takeaways from the results is that pupil-vergence coupling does not appear to behave as a fixed bias that applies uniformly across all conditions. Instead, it is state-dependent: PSA slopes varied widely both between participants and within participants across near and far viewing tasks. In some cases, the slope even flipped direction, making it difficult to justify global or participant-specific correction factors assumed to hold across depth states. While pupil size is influenced by factors such as cognitive state and arousal, near and far viewing also engage coordinated oculomotor mechanisms involving pupil size, accommodation, and vergence as part of the near triad. PSA slopes should therefore not be assumed to reflect purely random noise. Vergence is known to be misestimated by video-based eye tracking in a pupil-dependent manner, but the present results further suggest that the structure of this measurement error depends on the current depth state. 

% \vspace{0.5em}
\noindent\textbf{Constraining pupil dynamics reduces variability, but does not eliminate coupling.}
Introducing background luminance modulation reduced the spread of PSA slopes across participants. Under static illumination, slopes covered a wide range, whereas under modulated illumination they clustered much closer to zero. Because the near target was present in both conditions, this reduction is unlikely to be driven solely by target appearance or depth cues. One interpretation is that background luminance modulation constrains pupil size variability, reducing one source of instability in pupil-dependent vergence estimates. Under static lighting, slow and idiosyncratic pupil fluctuations leave slope estimates poorly constrained, whereas luminance modulation externally drives pupil dynamics while vergence remains relatively stable, causing fitted slopes to cluster toward zero. This does not remove pupil-related bias, though. Instead, it limits its magnitude. Even with modulation, depth-dependent differences remain, showing that pupil-related bias is reduced, not gone.

% \vspace{0.5em}
\noindent\textbf{Coupling persists without visual cues.}
The audio-cued control task supports this interpretation. Here, depth switching was driven by instruction, with no visual onset or offset of the near target and no changes in background illumination. Despite this, pupil-vergence coupling remained present and differed between conditions. This makes explanations that rely solely on visual transients less likely, such as stimulus appearance, disappearance, or contrast changes. Instead, the results suggest that the act of switching fixation depth appears sufficient to engage pupil-vergence interactions. Even when the visual scene is static, planning and executing vergence movements appears to be associated with pupil-related influences on measured vergence. 

% \vspace{0.5em}
\noindent\textbf{Implications for eye tracking researchers.}
Overall, these findings show that vergence estimates from video-based eye trackers remain usable for distinguishing depth states, but this does not imply accurate absolute vergence. Rather, vergence was sufficient for separating two controlled depth states, while still being systematically influenced by pupil dynamics and therefore not interpretable as a geometric signal. This applies even to modern, deep learning-based systems (e.g. Neon). Pupil-related influences persist, vary with viewing depth, and depend on task context, meaning that unmodeled pupil effects may introduce depth-dependent measurement bias that varies across users and task conditions. Importantly, constraining pupil dynamics, e.g. through luminance modulation, reduces variability in vergence estimates but does not eliminate pupil-related bias. Smaller or more consistent PSA slopes should therefore not be interpreted as evidence that the issue has been resolved. Static viewing conditions are not inherently ``clean,'' and depth switches alone can be sufficient to expose pupil-dependent instability. This suggests that experimental design and analysis choices, such as calibration strategy and window selection, can affect vergence-based conclusions and should explicitly account for pupil dynamics. \citeauthor{hooge2019pupil}~[\citeyear{hooge2019pupil}] showed that pupil-dependent vergence distortions in video-based eye tracking persist even during fixation at a single depth and are not fully removable with practical correction, with residual errors that can remain substantial, underscoring the need to control pupil dynamics and interpret vergence cautiously. Our results suggest that this limitation extends to physically grounded near-far viewing, and that reducing reliance on pupil features at the hardware level is a promising direction for mitigating pupil-related vergence bias.

% \vspace{0.5em}
\noindent\textbf{Limitations and Future Work.}
This work examines how pupil size influences vergence estimates under controlled near–far viewing and is not a validation of absolute vergence accuracy. The sample size is small ($N=8$), and analyses are conducted within participants to characterize measurement behavior rather than estimate population-level effects. Accordingly, findings reflect the presence and variability of pupil-vergence coupling under controlled conditions, not population-level effect sizes.A related constraint is the use of a single video-based eye tracker (Pupil Labs Neon). The study examines how pupil dynamics interact with vergence-based estimates rather than benchmarking hardware. Prior work shows that pupil-size artifacts arise from shifts in the apparent pupil center during dilation, producing systematic gaze deviations in video-based eye trackers \cite{drewes2014smaller}. Comparative studies further demonstrate that this effect occurs across modern video-based eye tracking systems, including Tobii Pro Glasses 2, Eyelink 1000 Plus, and Pupil Neon \cite{salari2025effect}, suggesting a general image-based measurement effect rather than a device-specific artifact. Because calibration and gaze-mapping differ across systems, the same underlying effect may vary in magnitude. Including multiple devices would therefore primarily reflect device-specific calibration rather than clarify the underlying measurement interaction. Finally, the experimental setup prioritizes control over ecological complexity. The beamsplitter enables physically grounded depth manipulation but does not cover more complex scenarios (e.g., dynamic scenes or large gaze angles). Similarly, luminance modulation constrains pupil dynamics but does not isolate causal mechanisms, and the observed coupling may reflect both oculomotor coordination and estimation bias.

While this study focuses on a single video-based eye tracker to isolate measurement behavior, future work will extend the analysis to alternative sensing modalities such as LFI-based eye tracking~\cite{meyer2025ambient}, which do not rely on image-based pupil center estimation, to test whether the observed effects generalize.

% \vspace{0.5em}
\noindent\textbf{Privacy and societal impact statement.}
This work analyzes eye-tracking measurement artifacts under controlled conditions and does not introduce additional privacy or societal risks beyond standard ethical considerations. Ethics approval was obtained from the Ethics Committee of the Technical University of Munich. 

\section{Conclusion}
We evaluated pupil-vergence coupling during physically grounded near-far viewing using a beamsplitter setup and a head-mounted eye tracker. Across depth switching, sustained fixation, and an audio-cued control without visual transients, vergence consistently separated depth states, while pupil size remained systematically related to vergence estimates with strong variability across participants and depth states. Luminance modulation reduced inter-participant spread in coupling estimates, yet the effect persisted, suggesting that smaller slopes do not imply the absence of pupil-related bias. Overall, vergence from video-based eye trackers provides a meaningful depth signal, but its interpretation cannot be treated as purely geometric without explicitly accounting for pupil dynamics and viewing context.

\begin{acks}
The project is supported by the Chips Joint Undertaking (Chips JU) and its members, including top-up funding by Denmark, Germany, Netherlands, Sweden, under grant agreement No. 101139942.
\end{acks}

\bibliographystyle{ACM-Reference-Format}
\bibliography{sample-base}

@article{ripps1962effect,
  title={The effect of pupil size on accommodation, convergence, and the AC/A ratio},
  author={Ripps, Harris and Chin, Newton B and Siegel, Irwin M and Breinin, Goodwin M},
  journal={Investigative ophthalmology \& visual science},
  volume={1},
  number={1},
  pages={127--135},
  year={1962},
  publisher={The Association for Research in Vision and Ophthalmology}
}

@article{drewes2014smaller,
  title={Smaller is better: Drift in gaze measurements due to pupil dynamics},
  author={Drewes, Jan and Zhu, Weina and Hu, Yingzhou and Hu, Xintian},
  journal={PloS one},
  volume={9},
  number={10},
  pages={e111197},
  year={2014},
  publisher={Public Library of Science San Francisco, USA}
}

@misc{balaban2018patterns,
  title={Patterns of pupillary activity during binocular disparity resolution. Front. Neurol. 9 (2018)},
  author={Balaban, CD and Kiderman, A and Szczupak, M and Ashmore, RC and Hoffer, ME},
  year={2018}
}

@article{jaschinski2016pupil,
  title={Pupil size affects measures of eye position in video eye tracking: implications for recording vergence accuracy},
  author={Jaschinski, Wolfgang},
  journal={Journal of Eye Movement Research},
  volume={9},
  number={4},
  year={2016}
}

@article{kahnemann1966pupil,
  title={Pupil diameter and load on memory},
  author={Kahnemann, D and Beatty, J},
  journal={Science},
  volume={154},
  number={3756},
  pages={1583--1585},
  year={1966}
}

@article{mathot2018pupillometry,
  title={Pupillometry: Psychology, physiology, and function},
  author={Math{\^o}t, Sebastiaan},
  journal={Journal of cognition},
  volume={1},
  number={1},
  pages={16},
  year={2018}
}

@article{hooge2019pupil,
  title={Do pupil-based binocular video eye trackers reliably measure vergence?},
  author={Hooge, Ignace TC and Hessels, Roy S and Nystr{\"o}m, Marcus},
  journal={Vision Research},
  volume={156},
  pages={1--9},
  year={2019},
  publisher={Elsevier}
}

@inproceedings{duchowski2011measuring,
  title={Measuring gaze depth with an eye tracker during stereoscopic display},
  author={Duchowski, Andrew T and Pelfrey, Brandon and House, Donald H and Wang, Rui},
  booktitle={Proceedings of the ACM SIGGRAPH symposium on applied perception in graphics and visualization},
  pages={15--22},
  year={2011}
}

@inproceedings{lee2017estimating,
  title={Estimating gaze depth using multi-layer perceptron},
  author={Lee, Youngho and Shin, Choonsung and Plopski, Alexander and Itoh, Yuta and Piumsomboon, Thammathip and Dey, Arindam and Lee, Gun and Kim, Seungwon and Billinghurst, Mark},
  booktitle={2017 International Symposium on Ubiquitous Virtual Reality (ISUVR)},
  pages={26--29},
  year={2017},
  organization={IEEE}
}

@inproceedings{oney2020evaluation,
  title={Evaluation of gaze depth estimation from eye tracking in augmented reality},
  author={Oney, Seyda and Rodrigues, Nils and Becher, Michael and Ertl, Thomas and Reina, Guido and Sedlmair, Michael and Weiskopf, Daniel},
  booktitle={ACM Symposium on Eye Tracking Research and Applications},
  pages={1--5},
  year={2020}
}

@inproceedings{von2025cnn,
  title={CNN-based estimation of gaze distance in virtual reality using eye tracking and depth data},
  author={von Behren, Anna-Lena and Sauer, Yannick and Severitt, Bj{\"o}rn and Wahl, Siegfried},
  booktitle={Proceedings of the 2025 Symposium on Eye Tracking Research and Applications},
  pages={1--7},
  year={2025}
}

@article{wang2023foveated,
  title={Foveated rendering: A state-of-the-art survey},
  author={Wang, Lili and Shi, Xuehuai and Liu, Yi},
  journal={Computational visual media},
  volume={9},
  number={2},
  pages={195--228},
  year={2023},
  publisher={TUP}
}

@article{guenter2012foveated,
  title={Foveated 3D graphics},
  author={Guenter, Brian and Finch, Mark and Drucker, Steven and Tan, Desney and Snyder, John},
  journal={ACM transactions on Graphics (tOG)},
  volume={31},
  number={6},
  pages={1--10},
  year={2012},
  publisher={ACM New York, NY, USA}
}

@inproceedings{mantiuk2011gaze,
  title={Gaze-dependent depth-of-field effect rendering in virtual environments},
  author={Mantiuk, Rados{\l}aw and Bazyluk, Bartosz and Tomaszewska, Anna},
  booktitle={International Conference on Serious Games Development and Applications},
  pages={1--12},
  year={2011},
  organization={Springer}
}

@inproceedings{mauderer2014depth,
  title={Depth perception with gaze-contingent depth of field},
  author={Mauderer, Michael and Conte, Simone and Nacenta, Miguel A and Vishwanath, Dhanraj},
  booktitle={Proceedings of the SIGCHI Conference on Human Factors in Computing Systems},
  pages={217--226},
  year={2014}
}

@inproceedings{duchowski2014reducing,
  title={Reducing visual discomfort of 3D stereoscopic displays with gaze-contingent depth-of-field},
  author={Duchowski, Andrew T and House, Donald H and Gestring, Jordan and Wang, Rui I and Krejtz, Krzysztof and Krejtz, Izabela and Mantiuk, Rados{\l}aw and Bazyluk, Bartosz},
  booktitle={Proceedings of the acm symposium on applied perception},
  pages={39--46},
  year={2014}
}

@incollection{schweigert2019eyepointing,
  title={Eyepointing: A gaze-based selection technique},
  author={Schweigert, Robin and Schwind, Valentin and Mayer, Sven},
  booktitle={Proceedings of mensch und computer 2019},
  pages={719--723},
  year={2019}
}

@inproceedings{kumar2007eyepoint,
  title={Eyepoint: practical pointing and selection using gaze and keyboard},
  author={Kumar, Manu and Paepcke, Andreas and Winograd, Terry},
  booktitle={Proceedings of the SIGCHI conference on Human factors in computing systems},
  pages={421--430},
  year={2007}
}

@article{myers1990topology,
  title={Topology of the near response triad},
  author={Myers, Glenn A and Stark, Lawrence},
  journal={Ophthalmic and Physiological Optics},
  volume={10},
  number={2},
  pages={175--181},
  year={1990},
  publisher={Wiley Online Library}
}

@article{hung1984near,
  title={The near response: modeling, instrumentation, and clinical applications},
  author={Hung, George K and Semmlon, John L and Ciuffreda, Kenneth J},
  journal={IEEE transactions on biomedical engineering},
  number={12},
  pages={910--919},
  year={1984},
  publisher={IEEE}
}

@article{cho2024hybrid,
  title={A Hybrid Gaze Distance Estimation via Cross-Reference of Vergence and Depth},
  author={Cho, Dae-Yong and Kang, Min-Koo},
  journal={IEEE Access},
  year={2024},
  publisher={IEEE}
}

@inproceedings{duchowski2014comparing,
  title={Comparing estimated gaze depth in virtual and physical environments},
  author={Duchowski, Andrew T and House, Donald H and Gestring, Jordan and Congdon, Robert and {\'S}wirski, Lech and Dodgson, Neil A and Krejtz, Krzysztof and Krejtz, Izabela},
  booktitle={Proceedings of the Symposium on Eye Tracking Research and Applications},
  pages={103--110},
  year={2014}
}

@article{baumann2023neon,
  title={Neon accuracy test report},
  author={Baumann, Chris and Dierkes, Kai},
  journal={Pupil Labs},
  volume={10},
  year={2023}
}

@article{hansen2009eye,
  title={In the eye of the beholder: A survey of models for eyes and gaze},
  author={Hansen, Dan Witzner and Ji, Qiang},
  journal={IEEE transactions on pattern analysis and machine intelligence},
  volume={32},
  number={3},
  pages={478--500},
  year={2009},
  publisher={IEEE}
}

@article{collewijn1997trajectories,
  title={Trajectories of the human binocular fixation point during conjugate and non-conjugate gaze-shifts},
  author={Collewijn, Han and Erkelens, Casper J and Steinman, Robert M},
  journal={Vision research},
  volume={37},
  number={8},
  pages={1049--1069},
  year={1997},
  publisher={Elsevier}
}

@article{mathot2015new,
  title={New light on the mind’s eye: The pupillary light response as active vision},
  author={Math{\^o}t, Sebastiaan and Van der Stigchel, Stefan},
  journal={Current directions in psychological science},
  volume={24},
  number={5},
  pages={374--378},
  year={2015},
  publisher={Sage Publications Sage CA: Los Angeles, CA}
}

@article{thaler2013best,
  title={What is the best fixation target? The effect of target shape on stability of fixational eye movements},
  author={Thaler, Lore and Sch{\"u}tz, Alexander C and Goodale, Melvyn A and Gegenfurtner, Karl R},
  journal={Vision research},
  volume={76},
  pages={31--42},
  year={2013},
  publisher={Elsevier}
}

@misc{robinson1966mechanics,
  title={The mechanics of human vergence eye movement},
  author={Robinson, DA},
  journal={Journal of Pediatric Ophthalmology \& Strabismus},
  volume={3},
  number={3},
  pages={31--37},
  year={1966},
  publisher={SLACK Incorporated Thorofare, NJ}
}

@article{hung1998dynamic,
  title={Dynamic model of the vergence eye movement system: simulations using MATLAB/SIMULINK},
  author={Hung, George K},
  journal={Computer methods and programs in biomedicine},
  volume={55},
  number={1},
  pages={59--68},
  year={1998},
  publisher={Elsevier}
}

@article{salari2025effect,
  title={The effect of pupil size on data quality in head-mounted eye trackers},
  author={Salari, Mohammadhossein and Niehorster, Diederick C and Nystr{\"o}m, Marcus and Bednarik, Roman},
  journal={Behavior Research Methods},
  volume={58},
  number={1},
  pages={17},
  year={2025},
  publisher={Springer}
}

@article{meyer2025ambient,
  title={Ambient Light Robust Eye-Tracking for Smart Glasses Using Laser Feedback Interferometry Sensors with Elongated Laser Beams},
  author={Meyer, Johannes and Zimmer, Alexander and Vilches, Sergio},
  journal={Proceedings of the ACM on Human-Computer Interaction},
  volume={9},
  number={3},
  pages={1--17},
  year={2025},
  publisher={ACM New York, NY, USA}
}

% \appendix
% \section*{Appendix}
% \section{Normality Analysis}
% \label{sec:normal}

\end{document}